\documentclass[english,aps,prl,showpacs,amsmath,amssymb,twocolumn]{revtex4}
\usepackage[ngerman]{babel}
\usepackage{graphicx}
\usepackage{bm}
\usepackage[latin1]{inputenc}

\begin{document}
\title{The uncertainty principle in terms of isoperimetric inequalities}
\author{Thomas Sch\"urmann}
\email{<t.schurmann@icloud.com>}
\affiliation{40223 D\"usseldorf, Germany}

\begin{abstract}
Simultaneous measurements of position and momentum are considered in $n$ dimensions. We find, that for a particle whose position is strictly localized in a compact domain $D\subset \mathbb{R}^n$ (spatial uncertainty) with non-empty boundary, the standard deviation of its momentum is sharply bounded by $\sigma_p \geq \lambda_1^{1/2}\hbar$, while $\lambda_1$ is the first Dirichlet eigenvalue of the Laplacian on $D$.
\end{abstract}

\maketitle

The most familiar formalization of the uncertainty principle of position and momentum is in terms of standard deviations
\cite{H27}\cite{H30}\cite{K27}
\begin{eqnarray}\label{K}
\sigma_p\sigma_x\geq \hbar/2.
\end{eqnarray}

In the corresponding measurement process the standard deviation of the position $\sigma_x$ is measured for a sample of particles initially prepared in a state $\psi$. Subsequently, the standard deviation of the momentum $\sigma_p$ is measured for another sample of particles, which is also prepared in the same state $\psi$. Thus, the statistical errors are corresponding to ensembles of measurements of different but identical prepared systems.

An alternative interpretation of the Heisenberg principle for {\it simultaneous} measurements has been given recently \cite{SH09}. The corresponding measurement process is as follows: Whenever a particle is strictly localized in a finite interval of length $\Delta x>0$ with probability 1, then the standard deviation of its momentum satisfies the inequality
\begin{eqnarray}\label{SH}
\sigma_p\Delta x\geq\pi\hbar.
\end{eqnarray}

Provided the wave function $\psi$ of the system is sufficiently regular at the boundary of $\Delta x$, the standard deviation of the momentum will remain finite and (\ref{SH}) is simply proved by applying the Wirtinger inequality \cite{Wirt}\cite{O78}.

In mathematics the Wirtinger inequality is closely connected to the fact that the circle is uniquely characterized by the property that among all simple closed plane curves of given length $L$, the circle of circumference $L$ encloses maximum area \cite{O78}\cite{R1894}. This property is most succintly expressed in the {\it isoperimetric inequality} $L^2> 4 \pi A$ where $A$ is the area enclosed by a curve of length $L$, and where equality holds if and only if the curve is a circle \cite{R1894}\cite{F23}\cite{K25}.

There are also {\it isoperimetric inequalities} of mathematical physics. They are special cases of isoperimetric problems in which typically some physical quantity, usually represented by the eigenvalues of a differential equation, is shown to be extremal for a circular or spherical domain. Extensive discussions of such problems can be found in the book of P\'{o}lya and Szeg\"{o} \cite{PS51} and the review article by Payne \cite{P67}.

The purpose of the present note is to establish the link between the measurement process of (\ref{SH}) and the corresponding analytic inequalities closely connected to the isoperimetric inequalities, such as Wirtinger's or Poincar\'{e}'s inequalities \cite{R1894}\cite{F23}\cite{K25}\cite{PS51}\cite{P67}\cite{KS84}\cite{K94}.

More precisely, let us consider non-relativistic particles in $n\geq 2$ spatial dimensions. In analogy to the 1-dimensional interval $\Delta x$ of (\ref{SH}), let $D\subset \mathbb{R}^n$ be a simply connected domain (the spatial uncertainty) with compact closure and (piecewise) smooth boundary $\partial D\neq\emptyset$. Then, the Hilbert basis of $L^2(D)$, the space of square-integrable functions on $D$, is defined by the Laplacian on $D$ with Dirichlet boundary conditions:
\begin{eqnarray}\label{Lap}
\Delta u + \lambda\,u&=&0\qquad\text{in }D,\\
u &=& 0\qquad\text{on } \partial D.
\end{eqnarray}

Let $\{\lambda_i\}$ be the set of eigenvalues and $\{u_i\}$ the orthonormal basis of eigenfunctions, $i=1,2,..$. It is well known that there are infinite many eigenvalues with no accumulation point: $0<\lambda_1\leq\lambda_2\leq...$ and $\lambda_i\to\infty$ as $i\to\infty$. The scalar product in $L^2(D)$ will be denoted by angular brackets, that is to write $\langle \phi|\psi\rangle$ for two state vectors $\phi,\psi\in L^2(D)$. Accordingly, the norm of $\psi$ is given by $||\psi||\equiv \sqrt{\langle \psi|\psi\rangle}$.

Now, we consider the standard deviation $\sigma_p$ of the momentum in the domain $D$. For every wave function $\psi\in L^2(D)$, the eigenvalue problem (\ref{Lap}) is the same for its real part and its imaginary part. Both are collinear and thus we only have to consider the real valued problem. In this case, it is easy to show by partial integrations that the mean value of the momentum operator $\hat{p}=-i\hbar\nabla$ is equal to zero and the standard deviation of the momentum is given by
\begin{eqnarray}\label{std1}
\sigma_p^2 = \hbar^2 ||\nabla \psi ||^2.
\end{eqnarray}
A sharp lower bound of $\sigma_p$ is now obtained by the associated variational characterization
\begin{eqnarray}\label{Ray}
\inf_{\psi\in L^2(D)\setminus\{0\}} \frac{||\nabla \psi ||^2}{||\psi ||^2 }=\lambda_1(D),
\end{eqnarray}
while the quotient on the left hand side is the well known Rayleigh quotient \cite{O78}\cite{R1894}. The right hand side is the first eigenvalue $\lambda_1$ of the Dirichlet Laplacian which is in general dependent on the shape of the domain \cite{O78}. After substitution of (\ref{Ray}) into (\ref{std1}), we obtain the corresponding inequality
\begin{eqnarray}\label{UP}
\sigma_p \geq \lambda_1^\frac{1}{2}\hbar\,.
\end{eqnarray}

That is, whenever there is a particle in a given domain $D$ with probability 1, then the standard deviation of the momentum is bounded by (\ref{UP}).\\

For an illustration, let us consider the case $n=2$. Then, the eigenvalue $\lambda_1$ is proportional to the square of the eigenfrequencies of an elastic, homogeneous, vibrating membrane with fixed boundary. The Rayleigh-Faber-Krahn inequality for the membrane (i.e. $n=2$) states that
\begin{eqnarray}\label{DFK2}
\lambda_1\geq\frac{\pi j_{0,1}^2}{A},
\end{eqnarray}
where $j_{0,1}$ is the first zero of the Besselfunction of order zero, and $A$ is the area of the membrane. Equality is attained in (\ref{DFK2}) only if the membrane is circular \cite{R1894}\cite{F23}\cite{K25}.

More general, the corresponding isoperimetric inequality in dimension $n$,
\begin{eqnarray}\label{Kn}
\lambda_1(D)\geq\left(\frac{C_n}{|D|}\right)^{2/n} j_{n/2-1,1}^2,
\end{eqnarray}
was proven by Krahn \cite{K94}. The expression $j_{m,1}$ is the first positive zero of the Besselfunction $J_m$, $|D|$ is
the volume of the domain and $C_n=\pi^{n/2}/\Gamma(n/2-1)$ is the volume of the $n$-dimensional unit ball. Equality is attained in (\ref{Kn}) if and only if $D$ is a ball. Let $d$ be the diameter of the $n$-dimensional ball, then we obtain the general inequality
\begin{eqnarray}\label{UP2}
\sigma_p\,d\geq 2 j_{n/2-1,1}\hbar.
\end{eqnarray}
\textbf{Proposition.} For dimension $n=1,2,3$, we get the following uncertainty relations:
\begin{eqnarray}\label{Bessel}
\sigma_p\,d&\geq& \pi\hbar\label{u1}\\
\sigma_p\,d&\geq& 4.8\hbar\label{u2}\\
\sigma_p\,d&\geq& 2\pi\hbar.\label{u3}
\end{eqnarray}
\textbf{Proof.} By applying (\ref{UP2}) for $n=1,2,3$. For the Bessel-zero with $n=1$ we have $j_{-1/2,1}=\pi/2$, for $n=2$ we have $j_{0,1}=2.40482555769...$ and for $n=3$ we get $j_{1/2,1}=\pi$.\\ 
\\
The first of these inequalities is equivalent to (\ref{SH}) for $\Delta x=d$, as it should be expected in one dimension. For the second inequality, we have applied the numerical approximation $j_{0,1}\approx 2.40$.

Actually, the derivation of (\ref{u1})-(\ref{u3}) is based on the assumption that the Hilbert space is considered with respect to the ordinary Euclidean position space. For a formal extension to general (curved) position spaces the Laplacian of the variational problem (\ref{Ray}) might be replaced by the corresponding Laplace-Beltrami operator of a Riemannian manifold. This procedure then leads to inequalities whose lower bound (in addition) depends on the Ricci curvature of the manifold. However, first of all it seems to be appropriate to ensure that the quantum mechanical measurement process corresponding to the generalized momentum operator is well defined. Otherwise, there is no way for experimental verification at all.

\acknowledgments


\begin{thebibliography}{99}
\bibitem{Wirt} This inequality states that for any bounded interval $Q=[a,b]$ and any continuously differentiable $f(x)\in C^1(Q)$ with $f(a)=f(b)=0$: $\int_Q |f|^2 dx\leq [(b-a)^2/\pi^2]\int_Q |f'|^2 dx$. The constant $(b-a)^2/\pi^2$ cannot be improved \cite{O78}.
\end{thebibliography}
\end{document}